# Verifying Computations with Streaming Interactive Proofs

Graham Cormode
AT&T Labs—Research
graham@research.att.com

Justin Thaler [*]
SEAS, Harvard University
jthaler@seas.harvard.edu

Ke Yi [†]
HKUST, Hong Kong
yike@cse.ust.hk

## ABSTRACT

When computation is outsourced, the data owner would like to be assured that the desired computation has been performed correctly by the service provider. In theory, proof systems can give the necessary assurance, but prior work is not sufficiently scalable or practical. In this paper, we develop new proof protocols for verifying computations which are streaming in nature: the verifier (data owner) needs only logarithmic space and a single pass over the input, and after observing the input follows a simple protocol with a prover (service provider) that takes logarithmic communication spread over a logarithmic number of rounds. These ensure that the computation is performed correctly: that the service provider has not made any errors or missed out some data. The guarantee is very strong: even if the service provider deliberately tries to cheat, there is only vanishingly small probability of doing so undetected, while a correct computation is always accepted.

We first observe that some theoretical results can be modified to work with streaming verifiers, showing that there are efficient protocols for problems in the complexity classes NP and NC. Our main results then seek to bridge the gap between theory and practice by developing usable protocols for a variety of problems of central importance in streaming and database processing. All these problems require linear space in the traditional streaming model, and therefore our protocols demonstrate that adding a prover can exponentially reduce the effort needed by the verifier. Our experimental results show that our protocols are practical and scalable.

## 1. INTRODUCTION

Efficient verification of computations has long played a central role in computer science. For example, the class of problems NP can be defined as the set of languages with certificates of membership that can be verified in polynomial time [2]. The most general verification model is the *interactive proof system* where there is a resource-limited *verifier* $\mathcal{V}$ and a more powerful *prover* $\mathcal{P}$ [2, Chapter 8]. To solve a problem, the verifier initiates a conversation with the prover, who solves the problem and *proves* the validity of his answer, following an established (randomized) protocol.

This model can be applied to the setting of *outsourcing computations* to a service provider. A wide variety of scenarios fit this template: in one extreme, a large business outsources its data to another company to store and process; at the other end of the scale, a hardware co-processor performs some computations within an embedded system. Over large data, the possibility for error increases: events like disk failure and memory read errors, which are usually thought unlikely, actually become quite common. A service provider who is paid for computation also has an economic incentive to take shortcuts, by returning an approximate result or only processing a sample of the data rather than the full amount. Hence, in these situations the data owner (the verifier in our model) wants to be assured that the computations performed by the service provider (the prover) are correct and complete, without having to take the effort to perform the computation himself. A natural approach is to use a proof protocol to *prove* the correctness of the answer. However, existing protocols for reliable delegation in complexity theory have so far been of theoretical interest: to our knowledge there have been no efforts to implement and use them. In part, this is because they require a lot of time and space for both parties. Historically, protocols have required the verifier to retain the full input, whereas in many practical situations the verifier cannot afford to do this, and instead outsources the storage of the data, often incrementally as updates are seen.

In this paper we introduce a proof system over data streams. That is, the verifier sees a data stream and tries to solve a (potentially difficult) problem with the help of a more powerful prover who sees the same stream. At the end of the stream, they conduct a conversation following an established protocol, through which an honest prover will always convince the verifier to accept its results, whereas any dishonest prover will not be able to fool the verifier into accepting a wrong answer with more than tiny probability.

Our work is motivated by developing applications in data outsourcing and trustworthy computing in general. In the increasingly popular model of "cloud computing", individuals or small businesses delegate the storage and processing of their data to a more powerful third party, the "cloud". This results in cost-savings, since the data owner no longer has to maintain expensive data storage and processing infrastructure. However, it is important that the data owner is fully assured that their data is processed accurately and completely by the cloud. In this paper, we provide protocols which allow the cloud to demonstrate that the results of queries are correct while keeping the data owner's computational effort minimal.

Our protocols only need the data owner (taking the role of verifier $\mathcal{V}$) to make a single streaming pass over the original data. This fits the cloud setting well: the pass over the input can take place incrementally as the verifier uploads data to the cloud. So the verifier never needs to hold the entirety of the data, since it can be shipped up to the cloud to store as it is collected. Without these new protocols, the verifier would either need to store the data in full, or retrieve the whole data from the cloud for each query: either way negates the benefits of the cloud model. Instead, our methods require the verifier to track only a logarithmic amount of information and follow a simple protocol with logarithmic communication to verify each query. Moreover, our results are of interest even when the verifier is able to store the entire input: they offer very lightweight and powerful techniques to verify computation, which happen to work in a streaming setting.

---

[*]Supported by a DoD NDSEG Fellowship, and partially by NSF grant CNS-0721491.

[†]Ke Yi is supported by a DAG and an RPC grant from HKUST.

*Motivating Example.* For concreteness, consider the motivating example of a cloud computing service which implements a key-value store. That is, the data owner sends (key, value) pairs to the cloud to be stored, intermingled with queries to retrieve the value associated with a particular key. For example, Dynamo supports two basic operations: *get* and *put* on key, value pairs [9]. In this scenario, the data owner never actually stores all the data at the same time (this is delegated to the cloud), but does see each piece as it is uploaded, one at a time: so we can think of this as giving a stream of (key, value) pairs. Our protocols allow the cloud to demonstrate that it has correctly retrieved the value of a key, as well as more complex operations, such as finding the next/previous key, finding the keys with large associated values, and computing aggregates over the key-value pairs (see Section 1.1 for definitions). □

Initial study in this area has identified the two critical parameters as the space used (by the verifier) and the total amount of communication between the two parties [6]. There are lower bounds which show that for many problems, the *product* of these two quantities must be at least linear in the size of the input when the verifier is not allowed to reply to the prover [6]. We use the notation of [6] and define an $(s,t)$-*protocol* to be one where the space usage of $\mathcal{V}$ is $O(s)$ and the total communication cost of the conversation between $\mathcal{P}$ and $\mathcal{V}$ is $O(t)$. We will measure both $s$ and $t$ in terms of words, where each word can represent quantities polynomial in $u$, the size of the universe over which the stream is defined. We additionally seek to minimize other quantities, such as the time costs of the prover and verifier, and the number of rounds of interaction.

Note that if $t = 0$ the model degenerates to the standard streaming model. We show that it is possible to drastically increase the computing power of the standard streaming model by allowing communication with a third party, verifiably solving many problems that are known to be hard in the standard streaming model.

We begin by observing that a key concept in proof systems, the *low-degree extension* of the input can be evaluated in a streaming fashion. Via prior results, this implies that (1) all problems in the complexity class NP have *computationally sound* protocols, so a dishonest prover cannot fool the verifier under standard cryptographic assumptions; and (2) all problems in NC have *statistically sound* protocols, meaning that the security guarantee holds even against computationally unbounded adversaries. These protocols have space and communication that is polynomial in the logarithm of the size of the input domain, $u$. These results can be contrasted with most results in the streaming literature, which normally apply only to one or a few problems at a time [21]. They demonstrate in principle the power of the streaming interactive proof model, but do not yield practical verification protocols.

Our main contributions in this paper are to provide protocols that are easy to implement and highly practical, for the following problems: *self join size*, *inner product*, *frequency moments*, *range query*, *range-sum query*, *dictionary*, *predecessor*, and *index*. These problems are all of considerable importance and all have been studied extensively in the standard streaming model and shown to require linear space [21]. As a result, approximations have to be allowed if sub-linear space is desired (for the first 3 problems); some of the problems do not have even approximate streaming algorithms (the last 5 problems). On the other hand, we solve them all *exactly* in our model. Our results are also asymptotically more efficient than those which would follow from the above theoretical results for NC problems. Formal definitions are in Section 1.1.

As well as requiring minimal space and communication for the verification, these new protocols are also very efficient in terms of both parties' running time. In particular, when processing the stream, the verifier spends $O(\log u)$ time per element. During verification the verifier spends $O(\log u)$ time while the (honest) prover runs in near-linear time. Thus, while our protocols are *secure* against a prover with unlimited power, an honest prover can execute our protocols efficiently. This makes our protocols simple enough to be deployed in real computation-outsourcing situations.

**Prior Work.** Cloud computing applications have also motivated a lengthy line of prior work in the cryptography community on "proofs of retrievability", which allow to verify that data is stored correctly by the cloud (see [16] and the references therein). In this paper, we provide "proofs of queries" which allow the cloud to demonstrate that the results of queries are correct while keeping the data owner's computational effort minimal.

Query verification/authentication for data outsourcing has been a popular topic recently in the database community. The majority of the work still requires the data owner to keep a full copy of the original data, e.g., [27]. More recently, there have been a few works which adopt a streaming-like model for the verifier, although they still require linear memory resources. For example, maintaining a Merkle tree [20] (a binary tree where each internal node is a cryptographic hash of its children) takes space linear in the size of the tree. Li et al. [19] considered verifying queries on a data stream with sliding windows via Merkle trees, hence the verifier's space is proportional to the window size. The protocol of Papadopoulos et al. [22] verifies a *continuous* query over streaming data, again requiring linear space on the verifier's side in the worst case.

Although interactive proof systems and other notions of verification have been extensively studied, they are mainly used to establish complexity results and hardness of approximation. Because they are usually concerned with answering "hard" problems, the (honest) prover's time cost is usually super-polynomial. Hence they have had little practical impact [26]. Recently, [14] reduced the cost of the prover to polynomial. Although of striking generality, the protocols that result are still complex, and require (polylogarithmically) many words of space and rounds of interaction. In contrast, our protocols for the problems defined in Section 1.1 require only logarithmic space and communication (and nearly linear running time for both prover and verifier). Thus we claim that they are practical for use in verifying outsourced query processing.

Our work is most directly motivated by prior work [28, 6] on verification of streaming computations that had stronger constraints. In the first model [28], the prover may send only the answer to the computation, which must be verified by $\mathcal{V}$ using a small sketch computed from the input stream of size $n$. Protocols were defined to verify identity and near-identity, and so because of the size of the answer, had small space ($s = 1$) and but large communication ($t = n$). Subsequent work showed that problems of showing a matching and connectedness in a graph could be solved in the same bounds, in a model where the prover's message was restricted to be a *permutation* of the input alone [23].

[6] introduced the notion of a streaming verifier, who must read first the input and then the proof under space constraints. They

26

allowed the prover to send a single message to the verifier, with no communication in the reverse direction. However, this does not dramatically improve the computational power. In this model, INDEX (see the definition in Section 1.1) can be solved using a $(\sqrt{n}, \sqrt{n})$-protocol and there is also a matching lower bound of $s \cdot t = \Omega(n)$ [6]; note that both $(n, 1)$- and $(1, n)$-protocols are easy, so the contribution of [6] is achieving a tradeoff between $s$ and $t$. In this paper, we show that allowing more interaction between the prover and the verifier exponentially reduces $s \cdot t$ for this and other problems that are hard in the standard streaming model.

## 1.1 Definitions and Problems

We first formally define a valid protocol:

DEFINITION 1. *Consider a prover $\mathcal{P}$ and verifier $\mathcal{V}$ who both observe a stream $\mathcal{A}$ and wish to compute a function $\beta(\mathcal{A})$. We assume $\mathcal{V}$ has access to a private random string $\mathcal{R}$, and one-way access to the input $\mathcal{A}$. After the stream is observed, $\mathcal{P}$ and $\mathcal{V}$ exchange a sequence of messages. Denote the output of $\mathcal{V}$ on input $\mathcal{A}$, given prover $\mathcal{P}$ and random string $\mathcal{R}$, by $out(\mathcal{V}, \mathcal{A}, \mathcal{R}, \mathcal{P})$. We allow $\mathcal{V}$ to output $\perp$ if $\mathcal{V}$ is not convinced that $\mathcal{P}$'s claim is valid.*

*Call $\mathcal{P}$ a valid prover with respect to $\mathcal{V}$ if for all streams $\mathcal{A}$, $\Pr_{\mathcal{R}}[out(\mathcal{V}, \mathcal{A}, \mathcal{R}, \mathcal{P}) = \beta(\mathcal{A})] = 1$. Call $\mathcal{V}$ a valid verifier for $\beta$ if*
*1. There exists at least one valid prover $\mathcal{P}$ with respect to $\mathcal{V}$.*
*2. For all provers $\mathcal{P}'$ and all streams $\mathcal{A}$,*

$$\Pr_{\mathcal{R}}[out(\mathcal{V}, \mathcal{A}, \mathcal{R}, \mathcal{P}') \notin \{\beta(\mathcal{A}), \perp\}] \leq 1/3.$$

Property 2 of Definition 1 defines *statistical soundness*. Notice the constant $\frac{1}{3}$ is arbitrary, and is chosen for consistency with standard definitions in complexity theory [2]. This should not be viewed as a limitation: note that as soon as we have such a prover, we can reduce probability of error to $p$, by repeating the protocol $O(\log 1/p)$ times in parallel, and rejecting if any rejects. In fact, our protocols let this probability be set arbitrarily small by appropriate choice of a parameter (the size of the finite field used), without needing to repeat the protocol.

DEFINITION 2. *We say the function $\beta$ possesses an r-round $(s, t)$ protocol, if there exists a valid verifier $\mathcal{V}$ for $\beta$ such that:*
*1. $\mathcal{V}$ has access to only $O(s)$ words of working memory.*
*2. There is a valid prover $\mathcal{P}$ for $\mathcal{V}$ such that $\mathcal{P}$ and $\mathcal{V}$ exchange at most $2r$ messages ($r$ messages in each direction), and the sum of the lengths of all messages is $O(t)$ words.*

We define some canonical problems to represent common queries on outsourced data, such as in a key-value store. Denote the universe from which data elements are drawn. as $[u] = \{0, \ldots, u - 1\}$.

INDEX: Given a stream of $u$ bits $b_1, \ldots, b_u$, followed by an index $q$, the answer is $b_q$.

DICTIONARY: The input is a stream of $n \leq u$ (key, value) pairs, where both the key and the value are drawn from the universe $[u]$, and all keys are distinct. The stream is followed by a query $q \in [u]$. If $q$ is one of the keys, then the answer is the corresponding value; otherwise the answer is "not found". This exactly captures the case of key-value stores such as Dynamo [9].

PREDECESSOR Given a stream of $n$ elements in $[u]$, followed by a query $q \in [u]$, the answer is the largest $p$ in the stream such that $p \leq q$. We assume that 0 always appears in the stream. SUCCESSOR is defined symmetrically. In a key-value store, this corresponds to finding the previous (next) key present relative to a query key.

RANGE QUERY: Given a stream of $n$ elements in $[u]$, followed by a range query $[q_L, q_R]$, the answer is the set of all elements in the stream between $q_L$ and $q_R$ inclusive.

RANGE-SUM: The input is a stream of $n$ (key, value) pairs, where both the key and the value are drawn from the universe $[u]$, and all keys are distinct. The stream is followed by a range query $[q_L, q_R]$. The answer is the sum of all the values with keys between $q_L$ and $q_R$ inclusive.

SELF-JOIN SIZE: Given a stream of $n$ elements from $[u]$, compute $\sum_{i \in [u]} a_i^2$ where $a_i$ is the number of occurrences of $i$ in the stream. This is also known as the *second frequency moment*.

FREQUENCY MOMENTS: In general, for any integer $k \geq 1$, $\sum_{i \in [u]} a_i^k$ is called the *k-th frequency moment* of the vector $\mathbf{a}$, written $F_k(\mathbf{a})$.

INNER PRODUCT (or JOIN SIZE): Given two streams $A$ and $B$ with frequency vectors $(a_1, \ldots, a_u)$ and $(b_1, \ldots, b_u)$, compute $\sum_{i \in [u]} a_i b_i$.

These queries are broken into two groups. The first four are *reporting* queries, which ask for elements from the input to be returned. INDEX is a classical problem that in the streaming model requires $\Omega(u)$ space [18]. It is clear that PREDECESSOR, DICTIONARY, RANGE QUERY, RANGE-SUM are all more general than INDEX and hence, also require linear space. These problems would be easy if the query were fixed before the data is seen. But in most applications, the user (the verifier) forms queries in response to other information that is only known after the data has arrived. For example, in database processing a typical range query may ask for all people in a given age range, where the range of interest is not known until after the database is instantiated.

The remaining queries are *aggregation* queries, computations that combine multiple elements from the input. SELF-JOIN SIZE requires linear space in the streaming model [1] to solve exactly (although there are space-efficient approximation algorithms). Since FREQUENCY MOMENTS and INNER PRODUCT are more general than SELF-JOIN SIZE, they also require linear space. In Section 6, we consider more general functions, such as heavy hitters, distinct elements ($F_0$), and frequency-based functions. These functions also require linear space to solve exactly, and certain functions like $F_{max}$ require polynomial space even to approximate [21]. These are additional functionalities that an advanced key-value store might support. For example, $F_0$ returns the number of distinct keys which are currently active, and the heavy hitters are the keys which have the largest values associated with them. These functions are also important in other contexts, e.g., tracking the heavy hitters over network data corresponds to the heaviest users or destinations [21].

**Outline.** In Section 2, we describe how existing proof systems can be modified to work with streaming verifiers, thereby providing space- and communication-efficient streaming protocols for all of NP and NC respectively. Subsequently, we improve upon these protocols for many problems of central importance in streaming and database processing. In Section 3 we give more efficient protocols to solve the aggregation queries (exactly), and in Section 4 we provide protocols for the reporting queries. In both cases, our protocols require only $O(\log u)$ space for the verifier $\mathcal{V}$, and $O(\log u)$ words of communication spread over $\log u$ rounds. An experimental study in Section 5 shows that these protocols are practical. In Section 6 we extend this approach to a class of *frequency-based functions*, providing protocols requiring $O(\log u)$ space and $\log u$ rounds, at the cost of more communication.

## 2. PROOFS AND STREAMS

We make use of a central concept from complexity theory, the *low-degree extension* (LDE) of the input, which is used in our protocols in the final step of checking. We explain how an LDE computation can be made over a stream of updates, and describe the immediate consequences for prior work which used the LDE.

**Input Model.** Each of the problems described in Section 1.1 above operates over an input stream. More generally, in all cases we can treat the input as defining an implicit vector $\mathbf{a}$, such that the value associated with key $i$ is the $i$th entry, $a_i$. The vector $\mathbf{a}$ has length $u$, which is typically too large to store (e.g. $u = 2^{128}$ if we consider the space of all possible IPv6 addresses). At the start of the stream, the vector $\mathbf{a} = (a_0, \ldots, a_{u-1})$ is initialized to $\mathbf{0}$. Each element in the stream is a pair of values, $(i, \delta)$ for integer $\delta$. A pair $(i, \delta)$ in the stream updates $a_i \leftarrow a_i + \delta$. This is a very general scenario: we can interpret pairs as adding to a value associated with each key (we allow negative values of $\delta$ to capture decrements or deletions). Or, if each $i$ occurs at most once in the stream, we can treat $(i, \delta)$ as associating the value $\delta$ with the key $i$.

**Low-Degree Extensions.** Given an input stream which defines a vector $\mathbf{a}$, we define a function $f_{\mathbf{a}}$ which is used in our protocols to check the prover's claims. Conceptually, we think of the vector $\mathbf{a}$ in terms of a *function* $f_{\mathbf{a}}$, so that $f_{\mathbf{a}}(i) = a_i$. By interpolation, $f_{\mathbf{a}}(i)$ can be represented as a polynomial, which is called the low-degree extension (LDE) of $\mathbf{a}$ [2]. LDEs are a standard tool in complexity theory. They give $\mathcal{V}$ surprising power to detect deviations by $\mathcal{P}$ from the prescribed protocol.

Given the LDE polynomial $f_{\mathbf{a}}$, we can also evaluate it at a location $r > u$. In our protocols, the verifier picks a secret location $r$ and computes $f_{\mathbf{a}}(r)$. In what follows, we formalize this notion, and explain how it is possible to compute $f_{\mathbf{a}}(r)$ in small space, incrementally as stream updates are seen.

First, we conceptually rearrange the data from a one-dimensional vector to a $d$ dimensional array. We let integer $\ell$ be a parameter, and assume for simplicity that $u = \ell^d$ is a power of $\ell$. Let $\mathbf{a} = (a_1, \ldots, a_u)$ be a vector in $[u]^u$. We first interpret $\mathbf{a}$ as a function $f'_{\mathbf{a}} : [\ell]^d \to [u]$ as follows: letting $(i)_k^\ell$ denote the $k$-th least significant digit of $i$ in base-$\ell$ representation, we associate each $i \in [u]$ with a vector $((i)_1^\ell, (i)_2^\ell, \ldots, (i)_d^\ell) \in [\ell]^d$, and define $f'_{\mathbf{a}}(i) = a_i$.

Pick a prime $p$ such that $u \leq p$. The *low-degree extension* (LDE) of $\mathbf{a}$ is a $d$-variate polynomial $f_{\mathbf{a}}$ over the field $\mathbb{Z}_p$ so that $f_{\mathbf{a}}(\mathbf{x}) = f'_{\mathbf{a}}(\mathbf{x})$ for all $\mathbf{x} \in [\ell]^d$. Notice since $f_{\mathbf{a}}$ is a polynomial over the field $\mathbb{Z}_p$, $f_{\mathbf{a}}(\mathbf{x})$ is defined for all $\mathbf{x} \in [p]^d$; $f_{\mathbf{a}}$ essentially extends the domain of $f'_{\mathbf{a}}$ from $[\ell]^d$ to $[p]^d$. Let $\mathbf{x} = (x_1, \ldots, x_d) \in [p]^d$ be a point in this $d$ dimensional space. The polynomial $f_{\mathbf{a}} : [p]^d \to \mathbb{Z}_p$ can be defined in terms of an indicator function $\chi_{\mathbf{v}}$ which is 1 at location $\mathbf{v} = (v_1, \ldots, v_d) \in [\ell]^d$ and zero elsewhere in $[\ell]^d$ via

$$\chi_{\mathbf{v}}(\mathbf{x}) = \prod_{j=1}^d \chi_{v_j}(x_j) \qquad (1)$$

where $\chi_k(x_j)$ is the Lagrange basis polynomial given by

$$\frac{(x_j - 0) \cdots (x_j - (k-1))(x_j - (k+1)) \cdots (x_j - (\ell-1))}{(k - 0) \cdots (k - (k-1))(k - (k+1)) \cdots (k - (\ell-1))}, \qquad (2)$$

which has the property that $\chi_k(x_j) = 1$ if $x_j = k$ and 0 for all $x_j \neq k$, $x_j \in [\ell]$. We then define $f_{\mathbf{a}}(\mathbf{x}) = \sum_{\mathbf{v} \in [\ell]^d} a_{\mathbf{v}} \chi_{\mathbf{v}}(\mathbf{x})$, which meets the requirement that $f_{\mathbf{a}}(\mathbf{x}) = f'_{\mathbf{a}}(\mathbf{x})$ when $\mathbf{x} \in [\ell]^d$.

**Streaming Computation of LDE.** We observe that while the polynomial $f_{\mathbf{a}}$ is defined over the very large domain $[p]^d$, it is actually very efficient to evaluate $f_{\mathbf{a}}(\mathbf{r})$ for some $\mathbf{r} \in [p]^d$ even when the input $\mathbf{a}$ is defined incrementally by a stream as in our input model. This follows from substituting $\mathbf{r}$ into (1): we obtain

$$f_{\mathbf{a}}(\mathbf{r}) = \sum_{\mathbf{v} \in [\ell]^d} a_{\mathbf{v}} \chi_{\mathbf{v}}(\mathbf{r}). \qquad (3)$$

Now observe that for fixed $\mathbf{r}$ this is a *linear* function of $\mathbf{a}$: a sum of multiples of the entries $a_{\mathbf{v}}$. So to compute $f_{\mathbf{a}}(\mathbf{r})$ in a streaming fashion, we can initialize $f_{\mathbf{0}}(\mathbf{r}) = 0$, and process each update $(i, \delta)$:

$$f_{\mathbf{a}}(\mathbf{r}) \leftarrow f_{\mathbf{a}}(\mathbf{r}) + \delta \chi_{\mathbf{v}(i)}(\mathbf{r}) \qquad (4)$$

where $\mathbf{v}(i)$ denotes the (canonical) remapping of $i$ into $[\ell]^d$. Note that $\chi_{\mathbf{v}}(\mathbf{r})$ can be computed in (at most) $O(d\ell)$ field operations, via (2); and $\mathcal{V}$ only needs to keep $f_{\mathbf{a}}(\mathbf{r})$ and $\mathbf{r}$, which takes $d+1$ words in $[p]$. Hence, we conclude

THEOREM 1. *The LDE $f_{\mathbf{a}}(\mathbf{r})$ can be computed over a stream of updates using space $O(d)$ and time per update $O(\ell d)$.*

**Initial Results.** We now describe results which follow by combining the streaming computation of LDE with prior results. Detailed analysis is in Appendix A. The constructions of [14] (respectively, [17]) yield small-space *non*-streaming verifiers and polylogarithmic communication for all problems in log-space uniform NC (respectively, NP), and achieve statistical (respectively, computational) soundness. The following theorems imply that both constructions can be implemented with a streaming verifier.

THEOREM 2. *There are computationally sound $(\text{poly}\log u, \log u)$ protocols for any problem in NP.*

Although Theorem 2 provides protocols with small space and communication, this does not yield a practical proof system. Even ignoring the complexity of constructing a PCP, the prover in a Universal Argument may need to solve an NP-hard problem just to determine the correct answer. However, Theorem 2 does demonstrate that in principle it is possible to have extremely efficient verification systems with streaming verifiers even for problems that are computationally difficult in a *non*-streaming setting.

THEOREM 3 (EXTENDING THEOREM 3 IN [14]). *There are statistically sound ($\text{poly}\log u$, $\text{poly}\log u$) protocols for any problem in log-space uniform NC.*

Here, NC is the class of all problems decidable by circuits of polynomial size and polylogarithmic depth; equivalently, the class of problems decidable in polylogarithmic time on a parallel computer with a polynomial number of processors. This class includes, for example, many fundamental matrix problems (e.g. determinant, product, inverse), and graph problems (e.g. minimum spanning tree, shortest paths) (see [2, Chapter 6]). Despite its powerful generality, the protocol implied by Theorem 3 is not optimal for many important functions in streaming and database applications. The remainder of this paper obtains improved, practical protocols for the fundamental problems listed in Section 1.1.

## 3. INTERACTIVE PROOFS FOR AGGREGATION QUERIES

We describe a protocol for the aggregation queries with a quadratic improvement over that obtained from Theorem 3.

### 3.1 SELF-JOIN SIZE Queries

We first explain the case of SELF-JOIN SIZE, which is $F_2 = \sum_{i \in [u]} a_i^2$. In the SELF-JOIN SIZE problem we are promised $\delta = 1$ for all updates $(i, \delta)$, but our protocol works even if we allow any integer $\delta$, positive or negative. This generality is useful for other queries considered later.

As in Section 2, let integer $\ell \geq 2$ be a parameter to be determined. We assume that $u$ is a power of $\ell$ for ease of presentation. Pick prime $p$ so $u \leq p \leq 2u$ (by Bertrand's Postulate, such a $p$ always exists). We also assume that $p$ is chosen so that $F_2 = O(p)$, to keep the analysis simple. The protocol we propose is similar to sum-check protocols in interactive proofs (see [2, Chapter 8]); given any $d$-variate polynomial $g$ over $\mathbb{Z}_p$, a sum-check protocol allows a polynomial-time verifier $\mathcal{V}$ to compute $\sum_{\mathbf{z} \in H^d} g(\mathbf{z})$ for any $H \subseteq \mathbb{Z}_p$, as long as $\mathcal{V}$ can evaluate $g$ at a randomly-chosen location in polynomial time. A sum-check protocol requires $d$ rounds of interaction, and the length of the $i$'th message from $\mathcal{P}$ to $\mathcal{V}$ is equal to $\deg_i(g)$, the degree of $g$ in the $i$'th variable.

Let $\mathbf{a}^2$ denote the entry-wise square of $\mathbf{a}$. A natural first attempt at a protocol for $F_2$ is to apply a sum-check protocol to the LDE $f_{\mathbf{a}^2}$ of $\mathbf{a}^2$ i.e. $f_{\mathbf{a}^2} = \sum_{\mathbf{v} \in [l]^d} a_\mathbf{v}^2 \chi_\mathbf{v}$. However, a streaming verifier cannot evaluate $f_{\mathbf{a}^2}$ at a random location because $\mathbf{a}^2$ is not a linear transform of the input. The key observation we need is that a streaming verifier *can* work with a different polynomial of slightly higher degree that also agrees with $\mathbf{a}^2$ on $[\ell]^d$. Specifically, the polynomial $f_\mathbf{a}^2 = (\sum_{\mathbf{v} \in [\ell]^d} a_\mathbf{v} \chi_\mathbf{v})^2$. That is, $\mathcal{V}$ can evaluate the polynomial $f_\mathbf{a}^2$ at a random location $\mathbf{r}$: $\mathcal{V}$ computes $f_\mathbf{a}(\mathbf{r})$ as in Section 2, and uses the identity $f_\mathbf{a}^2(\mathbf{r}) = f_\mathbf{a}(\mathbf{r})^2$. We can then apply a sum-check protocol to $f_\mathbf{a}^2$ in our model; details follow.

**The protocol.** Before observing the stream, the verifier picks a random location $\mathbf{r} = (r_1, \ldots, r_d) \in [p]^d$. Both prover and verifier observe the stream which defines $\mathbf{a}$. The verifier $\mathcal{V}$ evaluates the LDE $f_\mathbf{a}(\mathbf{r})$ in incremental fashion, as described in Section 2.

After observing the stream, the verification protocol proceeds in $d$ rounds as follows. In the first round, the prover sends a polynomial $g_1(x_1)$, and claims that

$$g_1(x_1) = \sum_{x_2,\ldots,x_d \in [\ell]^{d-1}} f_\mathbf{a}^2(x_1, x_2, \ldots, x_d). \quad (5)$$

Observe that if $g_1$ is as claimed, then $F_2(\mathbf{a}) = \sum_{x_1 \in [\ell]} g_1(x_1)$.

Since the polynomial $g_1(x_1)$ has degree $2(\ell-1)$, it can be described in $2(\ell-1)+1$ words.

Then, in round $j > 1$, the verifier sends $r_{j-1}$ to the prover. In return, the prover sends a polynomial $g_j(x_j)$, and claims that

$$g_j(x_j) = \sum_{x_{j+1},\ldots,x_d \in [\ell]^{d-j}} f_\mathbf{a}^2(r_1,\ldots,r_{j-1},x_j,x_{j+1},\ldots,x_d). \quad (6)$$

The verifier compares the two most recent polynomials by checking

$$g_{j-1}(r_{j-1}) = \sum_{x_j \in [\ell]} g_j(x_j)$$

and rejecting otherwise. The verifier also rejects if the degree of $g$ is too high: each $g$ should have degree $2(\ell-1)$.

In the final round, the prover has sent $g_d$ which is claimed to be

$$g_d(x_d) = f_\mathbf{a}^2(r_1, \ldots, r_{d-1}, x_d)$$

The verifier can now check that $g_d(r_d) = f_\mathbf{a}^2(\mathbf{r})$ (recall that the verifier tracked $f_\mathbf{a}(\mathbf{r})$ incrementally in the stream). If this test succeeds (and so do all previous tests), then the verifier accepts, and is convinced that $F_2(\mathbf{a}) = \sum_{x_1 \in [\ell]} g_1(x_1)$. We defer the detailed proof of correctness and the analysis of the prover's cost to Appendix B.1.

**Analysis of space and communication.** The communication cost of the protocol is dominated by the polynomials being sent by the prover. Each polynomial can be sent in $O(\ell)$ words, so over the $d$ rounds, the total cost is $O(d\ell)$ communication. The space required by the verifier is bounded by having to remember $\mathbf{r}$, $f_\mathbf{a}(\mathbf{r})$ and a constant number of polynomials (the verifier can "forget" intermediate polynomials once they have been checked). The total cost of this is $O(d+\ell)$ words. Probably the most economical tradeoff is reached by picking $\ell = 2$ and $d = \log u$, yielding both communication and space cost for $\mathcal{V}$ of $O(\log u)$ words.[1] Combining these settings with Lemma 1 and the analysis in Appendix B.1, we have:

---

[1]It is possible to tradeoff smaller space for more communication by, say, setting $\ell = \log^\varepsilon u$ and $d = \frac{\log u}{\varepsilon \log \log u}$ for any small constant $\varepsilon > 0$, which yields a protocol with $O(\frac{\log u}{\log \log u})$ space and $O(\log^{1+\varepsilon} u)$ communication.

THEOREM 4. *There is a $(\log u, \log u)$-protocol for* SELF-JOIN SIZE *with probability of failure $O(\frac{\log u}{p})$. The prover's total time is $O(\min(u, n \log u/n))$; the verifier takes time $O(\log u)$ per update.*

*Remarks.* Lemma 1 in Appendix B.1 shows that the failure probability is $2\ell d/p = 4 \log u/p$. It can be made as low as $O(\frac{\log u}{u^c})$ for any constant $c$, by choosing $p$ larger than $u^c$, without changing the asymptotic bounds. Notice that the smallest-depth circuit computing $F_2$ has depth $\Theta(\log u)$, as any function that depends on all bits of the input requires at least logarithmic depth. Therefore, Theorem 3 yields a $(\log^2 u, \log^2 u)$-protocol for $F_2$, and our protocol represents a quadratic improvement in both parameters.

### 3.2 Other Problems

Our protocol for $F_2$ can be easily modified to support the other aggregation queries listed in Section 1.1.

**Higher frequency moments.** The protocol outlined above naturally extends to higher frequency moments, or the sum of any polynomial function of $a_i$. For example, we can simply replace $f_\mathbf{a}^2$ with $f_\mathbf{a}^k$ in (5) and (6) to compute the $k$-th frequency moment $F_k$ (again, assuming $u$ is chosen large enough so $F_k < u$). The communication cost increases to $O(k \log u)$, since each $g_j$ now has degree $O(k)$ and so requires correspondingly more words to describe. However, the verifier's space bound remains at $O(\log u)$ words.

**Inner product.** Given two streams defining two vectors $\mathbf{a}$ and $\mathbf{b}$, their inner product is defined by $\mathbf{a} \cdot \mathbf{b} = \sum_{i \in [u]} a_i b_i$. Observe that $F_2(\mathbf{a}+\mathbf{b}) = F_2(\mathbf{a}) + F_2(\mathbf{b}) + 2\mathbf{a} \cdot \mathbf{b}$. Hence, the inner product can be verified by verifying three $F_2$ computations.

More directly, the above protocol for $F_2$ can be adapted to verify the inner product: instead of providing polynomials which are claimed to be sums of $f_\mathbf{a}^2$, we now have two LDEs $f_\mathbf{a}$ and $f_\mathbf{b}$ which encode $\mathbf{a}$ and $\mathbf{b}$ respectively. The verifier again picks a random $\mathbf{r}$, and evaluates LDEs $f_\mathbf{a}(\mathbf{r})$ and $f_\mathbf{b}(\mathbf{r})$ over the stream. The prover now provides polynomials that are claimed to be sums of $f_\mathbf{a} f_\mathbf{b}$. This observation is useful for the RANGE-SUM problem.

**Range-sum.** It is easy to see that RANGE-SUM is a special case of INNER PRODUCT. Here, every (key, value) pair in the input stream can considered as updating $i =$key with $\delta =$value to generate $\mathbf{a}$. When the query $[q_L, q_R]$ is given, the verifier defines $b_{q_L} = \cdots = b_{q_R} = 1$ and $b_i = 0$ for all other $i$. One technical issue is that computing $f_\mathbf{b}(\mathbf{r})$ directly from the definition requires $O(u \log u)$ time. However, the verifier can compute it much faster for such $\mathbf{b}$. Again fix $\ell = 2$. Decompose the range $[q_L, q_R]$ into $O(\log u)$ canonical intervals where each interval consists of all locations $\mathbf{v}$ where $v_{j+1}, \ldots, v_d$ are fixed while all possible $(v_1, \ldots, v_j) \in [2]^j$ for some $j$ occur. The value of $f_\mathbf{b}(\mathbf{r})$ in each such interval is

$$f_\mathbf{b}(\mathbf{r}) = \sum_{(v_1,\ldots,v_j) \in [2]^j} \chi_{(v_1,\ldots,v_d)}(\mathbf{r})$$

$$= \sum_{(v_1,\ldots,v_j) \in [2]^j} \prod_{k=1}^j \chi_{v_k}(r_k) \cdot \prod_{k=j+1}^d \chi_{v_k}(r_k)$$

$$= \prod_{k=j+1}^d \chi_{v_k}(r_k) \cdot \left( \sum_{(v_1,\ldots,v_j) \in [2]^j} \prod_{k=1}^j \chi_{v_j}(r_j) \right)$$

$$= \prod_{k=j+1}^d \chi_{v_k}(r_k) \cdot \left( \prod_{k=1}^j \left( \chi_0(r_j) + \chi_1(r_j) \right) \right) = \prod_{k=j+1}^d \chi_{v_k}(r_k),$$

which can be computed in $O(\log u)$ time. The final evaluation is found by summing over the $O(\log u)$ canonical intervals, so the time to compute $f_\mathbf{b}(\mathbf{r})$ is $O(\log^2 u)$. This is used to determine whether $g_d(r_d) = f_\mathbf{a}(\mathbf{r}) f_\mathbf{b}(\mathbf{r})$. Hence, the verifier can continue the rest of the verification process in $O(\log u)$ rounds as before.

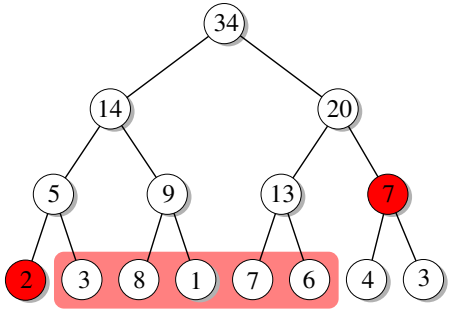

**Figure 1:** Example tree $\mathcal{T}$ over input vector $[2,3,8,1,7,6,4,3]$ and sub-vector query $(1,5)$.

# 4. INTERACTIVE PROOFS FOR REPORTING QUERIES

We first present an interactive proof protocol for a class of SUB-VECTOR queries, which is powerful enough to incorporate INDEX, DICTIONARY, PREDECESSOR, and RANGE QUERY as special cases.

## 4.1 SUB-VECTOR Queries

As before, the input is a stream of $n$ pairs $(i,\delta)$, which sets $a_i \leftarrow a_i + \delta$, defining a vector $\mathbf{a} = (a_1,\ldots,a_u)$ in $[u]^u$. The correct answer to a SUB-VECTOR query specified by a range $[q_L, q_R]$ is the $k$ nonzero entries in the sub-vector $(a_{q_L},\ldots,a_{q_R})$.

**The protocol.** Let $p$ be a prime such that $u < p \leq 2u$. The verifier $\mathcal{V}$ *conceptually* builds a tree $\mathcal{T}$ of constant degree $\ell$ on the vector $\mathbf{a}$. $\mathcal{V}$ first generates $\log u$ independent random numbers $r_1,\ldots,r_{\log_\ell u}$ uniformly from $[p]$. For simplicity, we describe the case for $\ell = 2$. For each node $v$ of the tree, we define a "hash" value as follows. For the $i$-th leaf $v$, set $v = a_i$. For an internal node $v$ at level $j$ (the leaves are at level 0), define

$$v = v_L + v_R r_j, \qquad (7)$$

where $v_L$ and $v_R$ are the left and right child of $v$, respectively. Additions and multiplications are done over the field $\mathbb{Z}_p$ as in Section 3. Denote the root of the tree by $t$. The verifier is only required to keep $r_1,\ldots,r_{\log u}$ and $t$. Later we show that $\mathcal{V}$ can compute $t$ without materializing the binary tree $\mathcal{T}$, and that this is essentially an LDE computation.

We first present the interactive verification protocol between $\mathcal{P}$ and $\mathcal{V}$ after the input has been observed by both. The verifier only needs $r_1,\ldots,r_{\log u}$, $t$, and the query range $[q_L, q_R]$ to carry out the protocol. First $\mathcal{V}$ sends $q_L$ and $q_R$ to $\mathcal{P}$, and $\mathcal{P}$ returns the claimed sub-vector, say, $a'_{q_L},\ldots,a'_{q_R}$ ($\mathcal{P}$ actually only needs to return the $k$ nonzero entries). In addition, if $q_L$ is even, $\mathcal{P}$ also returns $a'_{q_L-1}$; if $q_R$ is odd, $\mathcal{P}$ also returns $a'_{q_R+1}$. Then $\mathcal{V}$ tries to verify whether $a_i = a'_i$ for all $q_L \leq i \leq q_R$ using the following protocol. The general idea is to reconstruct $\mathcal{T}$ using information provided by $\mathcal{P}$. If $\mathcal{P}$ is behaving correctly, the (hash of the) reconstructed root, say $t'$, should be the same as $t$; otherwise with high probability $t' \neq t$ and $\mathcal{V}$ will reject. Define $\gamma^{(j)}(i)$ to be the ancestor of the $i$-th leaf of $\mathcal{T}$ on level $j$. The protocol proceeds in $\log u - 1$ rounds, and maintains the invariant that after the $j$-th round, $\mathcal{V}$ has reconstructed $\gamma^{(j+1)}(i)$ for all $q_L \leq i \leq q_R$. The invariant is easily established initially ($j = 0$) since $\mathcal{P}$ provides $a'_{q_L},\ldots,a'_{q_R}$ and the siblings of $a'_{q_L}$ and $a'_{q_R}$ if needed. In the $j$-th round, $\mathcal{V}$ sends $r_j$ to $\mathcal{P}$. Having $r_1,\ldots,r_j$ to hand, $\mathcal{P}$ can construct the $j$-th level of $\mathcal{T}$. $\mathcal{P}$ then returns to $\mathcal{V}$ the siblings of $\gamma^{(j)}(q_L)$ and $\gamma^{(j)}(q_R)$ if they are needed by $\mathcal{V}$. Then $\mathcal{V}$ reconstructs $\gamma^{(j+1)}(i)$ for all $q_L \leq i \leq q_R$. At the end of the $(\log u - 1)$-th round, $\mathcal{V}$ has reconstructed $\gamma^{(\log u)}(i) = t'$, and checks that $t = t'$. If so, then the initial answer provided by $\mathcal{P}$ is accepted, otherwise it is rejected.

**Example.** Figure 1 shows a small example on the vector $\mathbf{a} = [2,3,8,1,7,6,4,3]$. We fix the hash function parameters $r = [1,1,1]$ to keep the example simple (ordinarily these parameters are chosen randomly), and show the hash value inside each node. For the range $(2,6)$, in the first round the prover reports the sub-vector $[3,8,1,7,6]$ (shown highlighted). Since the left endpoint of this range is even, $\mathcal{P}$ also reports $a_1 = 2$. From this, $\mathcal{V}$ is able to compute some hashes at the next level: 5, 9 and 13. After sending $r_1$ to $\mathcal{P}$, $\mathcal{V}$ received the fact that the hash of the range $(7,8)$ is 7. From this, $\mathcal{V}$ can compute the final hash values and check that they agree with the precomputed hash value of $t$, 34. □

We prove the next theorem in Appendix B.2.

THEOREM 5. *There is a $(\log u, \log u + k)$-protocol for* SUB-VECTOR, *with failure probability $O(\frac{\log u}{p})$. The prover's total time is $O(\min(u, n\log u/n))$, the verifier takes time $O(\log u)$ per update.*

## 4.2 Answering Reporting Queries

We now show how to answer the reporting queries using the solution to SUB-VECTOR.

- It is straightforward to solve RANGE QUERY using SUB-VECTOR: each element $i$ in the stream is interpreted as a vector update with $\delta = 1$, and vector entries with non-zero counts intersecting the range give the required answer.

- INDEX can be interpreted as a special case of RANGE QUERY with $q_L = q_R = q$.

- For DICTIONARY, we must distinguish between "not found" and a value of 0. We do this by using a universe size of $[u+1]$ for the values: each value is incremented on insertion. At query time, if the retrieved value is 0, the result is "not found"; otherwise the value is decremented by 1 and returned.

- For PREDECESSOR, we interpret each key in the stream as an update with $\delta = 1$. In the protocol $\mathcal{V}$ first asks for the index of the predecessor of $q$, say $q'$, and then verifies that the sub-vector $(a_{q'},\ldots,a_q) = (1,0,\ldots,0)$, with communication cost $O(\log u)$ (since $k = 0$).

COROLLARY 1. *There is a $(\log u, \log u)$-protocol for* DICTIONARY, INDEX *and* PREDECESSOR *where the prover takes time $O(\min(u, n\log u/n))$. There is a $(\log u, \log(u) + k)$-protocol for* RANGE QUERY *where the prover's time is $O(k + \min(u, n\log u/n))$. For all protocols, the verifier takes time $O(\log u)$.*

# 5. EXPERIMENTAL STUDY

We performed a brief experimental study to validate our claims that the protocols described are practical. We compared protocols for both the reporting queries and aggregates queries. Specifically, we compared the multi-round protocols for $F_2$ described in Section 3 to the single round protocol given in [6], which can be seen as a protocol in our setting with $d = 2$ and $\ell = \sqrt{u}$. For reporting queries, we show the behavior of our SUB-VECTOR protocol, and we present experimental results when the length $q_R - q_L$ of the sub-vector queried is 1000. Together, these determine the performance of the 8 core queries: the three aggregate queries are based on the $F_2$ protocol, while the five reporting queries are based on the SUB-VECTOR protocol.

Our implementation was made in C++: it performed the computations of both parties, and measured the resources consumed by the protocols. All programs were compiled with g++ using the -O3 optimization flag. For the data, we generated synthetic streams with $u = n$ where the number of occurrences of each item $i$ was picked uniformly in the range $[0, 1000]$. Note that the choice of data does not affect the behavior of the protocols: their guarantees do not depend on the data, but rather on the random choices of the verifier. The computations were made over the field of size $p = 2^{61} - 1$, giving a probability of $4 \cdot 61/p \approx 10^{-16}$ of the verifier being fooled by a dishonest prover. These computations were executed using native 64-bit arithmetic, so increasing this probability is unlikely to affect performance. This probability could be reduced further to, e.g. $4 \cdot 127/(2^{127} - 1) < 10^{-35}$, at the cost of using 128 bit arithmetic.

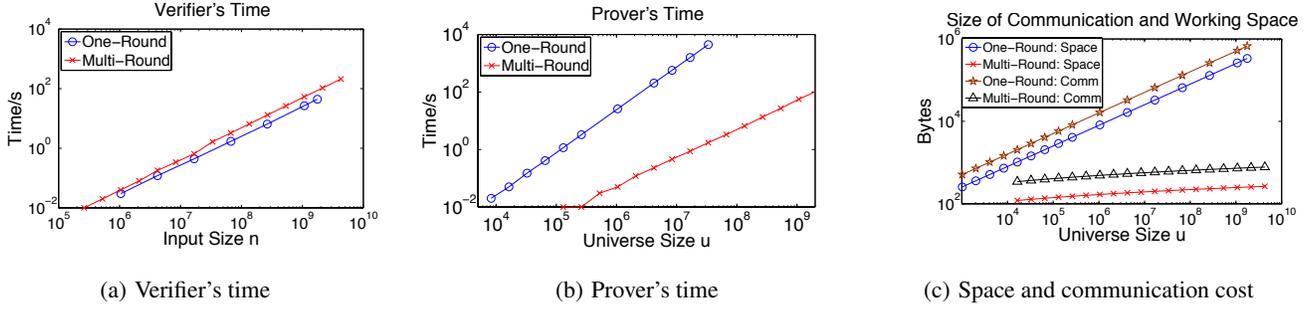

(a) Verifier's time  (b) Prover's time  (c) Space and communication cost

Figure 2: Experimental results for $F_2$

We evaluated the protocols on a single core of a multi-core machine with 64-bit AMD Opteron processors and 32 GB of memory available. The large memory let us experiment with universes of size several billion, with the prover able to store the entire frequency vector in memory. We measured all relevant costs: the time for $\mathcal{V}$ to compute the check information from the stream, for $\mathcal{P}$ to generate the proof, and for $\mathcal{V}$ to verify this proof. We also measured the space required by $\mathcal{V}$, and the size of the proof provided by $\mathcal{P}$.

**Experimental Results.** When the prover was honest, both protocols always accepted. We also tried modifying the prover's messages, by changing some pieces of the proof, or computing the proof for a slightly modified stream. In all cases, the protocols caught the error, and rejected the proof. We conclude that the protocols work as analyzed, and the focus of our experimental study is to understand how they scale to large volumes of data.

Figure 2 shows the behavior of the $F_2$ protocols as the data size varies. First, Figure 2(a) shows the time for $\mathcal{V}$ to process the stream to compute the necessary LDEs as the stream length increases. Both show a linear trend (here, plotted on log scale). Moreover, both take comparable time (within a constant factor), with the multi-round verifier processing about 21 million updates per second, and the single round $\mathcal{V}$ processing 35 million. The similarity is not surprising: both methods are taking each element of the stream and computing the product of the frequency with a function of the element's index $i$ and the random parameter $r$. The effort in computing this function is roughly similar in both cases. The single round $\mathcal{V}$ has a slight advantage, since it can compute and use lookup tables within the $O(\sqrt{u})$ space bound [6], while the multi-round verifier limited to logarithmic space must recompute some values multiple times. The time to check the proof is essentially negligible: less than a millisecond across all data sizes. Hence, we do not consider this a significant cost.

Figure 2(b) shows a clear separation between the two methods in $\mathcal{P}$'s effort in generating the proof. Here, we measure total time across all rounds in the multi-round case, and the time to generate the single round proof. The cost in the multi-round case is dramat-ically lower: it takes minutes to process inputs with $u = 2^{22}$ in the single round case, whereas the same data requires just 0.2 seconds when using the multi-round approach. Worse, this cost grows with $u^{3/2}$, as seen with the steeper line: doubling the input size increases the cost by a factor of 2.8. In contrast, the multi-round cost grew linearly with $u$. Across all values of $u$, the multi-round prover processed 20-21 million updates per second. Meanwhile, at $u = 2^{20}$, the single-round $\mathcal{P}$ processed roughly 40,000 updates per second, while at $u = 2^{24}$, $\mathcal{P}$ processed only 10,000. Thus the chief bottleneck of these protocols seems to be $\mathcal{P}$'s time to make the proof.

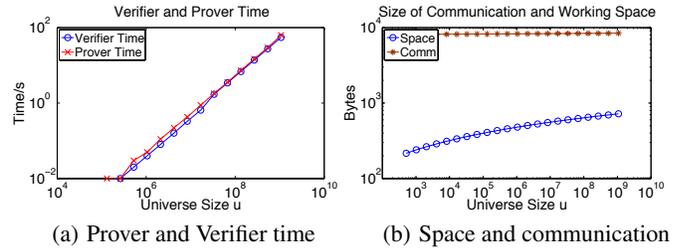

(a) Prover and Verifier time  (b) Space and communication

Figure 3: Experimental results for SUB-VECTOR

The trend is similar for the space resources required to execute the protocol. In the single round case, both the verifier's space and size of the proof grow proportional to $\sqrt{u}$. This is not impossibly large: Figure 2(c) shows that for $u$ of the order of 1 billion, both these quantities are comfortably under a megabyte. Nevertheless, it is still orders of magnitude larger than the sizes seen in the multi-round protocol: there, the space required and proof size are never more than 1KB even when handling gigabytes of data.

The results for reporting (SUB-VECTOR) queries are quite similar (Figure 3). Here, there are no comparable protocols for this query. The verifier's time is about the same as for the $F_2$ query: unsurprising, as in both protocols $\mathcal{V}$ evaluates the LDE of the input at a point $\mathbf{r}$. The prover's time is similarly fast, since the amount of work it has to do is about the same as the verifier (it has to compute hash values of various substrings of the input). The space cost of the verifier is minimal, primarily just to store $\mathbf{r}$ and some intermediate values. The communication cost is dominated by the cost of reporting the answer (1000 values): the rest is less than 1KB.

Our experiments focus on the case $u = n$. We can extrapolate the prover's cost, which scales as $O(\min(u, n \log u/n))$, to larger examples. Consider 1TB of IPv6 web addresses; this is approximately $6 \times 10^{10}$ IPv6 addresses, each drawn over a $\log u = 128$ bit domain. Figure 2(b) shows that processing $10^{10}$ updates from a domain of size $10^{10}$ takes approximately 500 seconds. In our IPv6

31

example, the input has 6 times more values, and the value of $\log u$ is approximately 4 times larger, so extrapolating we would expect our (uniprocessor) prover to take about 24 times longer to process this input, i.e. about 12,000 seconds (200 minutes). Note that this is comparable to the time to read this much data resident on disk [13].

In summary, the methods we have developed are applicable to genuinely large data sets, defined over a domain of size hundreds of millions to billions. Our implementation is capable of processing such datasets within a matter of seconds or minutes.

## 6. EXTENSIONS

We next consider how to treat other functions in the streaming interactive proof setting. We first consider some functions which are of interest in streaming, such as heavy hitters, and $k$-largest. We then discuss extensions of the framework to handle a more general class of "frequency-based functions".

### 6.1 Other Specific Functions

**Heavy Hitters.** The heavy hitters (HHs) are those items whose frequencies exceed a fraction $\phi$ of the total stream length $n$. In verifying the claimed set of HHs, $\mathcal{V}$ must ensure that all claimed HHs indeed have high enough frequency, and moreover no HHs are *omitted*. To convince $\mathcal{V}$ of this, $\mathcal{P}$ will combine a succinct witness set with a generalization of the SUB-VECTOR protocol to give a $(1/\phi \log u, 1/\phi \log u)$ protocol for verifying the heavy hitters and their frequencies. As in our SUB-VECTOR protocol, $\mathcal{V}$ conceptually builds a binary tree $\mathcal{T}$ with leaves corresponding to entries of **a**, and a random hash function associated with each level of $\mathcal{T}$. We augment each internal node $v$ with a third child $c_v$. $c_v$ is a leaf whose value is the sum of the frequencies of all *descendents* of $v$, the *subtree count* of $v$. The hash function now takes *three* arguments as input. It follows that $\mathcal{V}$ can still compute the hash $t$ of the root of this tree in logarithmic space, and $O(\log u)$ time per update.

In the $l$th round, the prover lists all leaves at level $l$ whose sub-tree count is at least $\phi n$, their siblings, as well as their hash value and their subtree counts (so the hash of their parent can be computed). In addition, $\mathcal{P}$ provides all leaves whose subtree count is less than $\phi n$ but whose parent has subtree count at least $\phi n$; these nodes serve as *witnesses* to the fact that none of their descendants are heavy hitters, enabling $\mathcal{V}$ to ensure that no heavy hitters are omitted. This procedure is repeated for each level of $\mathcal{T}$; note that for each node $v$ whose value $\mathcal{P}$ provides, all ancestors of $v$ and their siblings (i.e. all nodes on $v$'s "authentication path") are also provided, because the subtree count of any ancestor is at least as high as the subtree count of $v$. Therefore, $\mathcal{V}$ can compare the hash of the root (calculated while observing the stream) to the value provided by $\mathcal{P}$, and the proof of soundness is analogous to that for the SUB-VECTOR protocol.

In total, there are at most $O(1/\phi \log u)$ nodes provided by $\mathcal{P}$: for each level $l$, the sum of the sub-tree counts of nodes at level $l$ is $n$, and therefore there are $O(1/\phi)$ nodes at each level which have sub-tree count exceeding $\phi n$ or whose parent has subtree count exceeding $\phi n$. Hence, the size of the proof is at most $O(1/\phi \log u)$, and the time costs are as for the SUB-VECTOR protocol.

The protocol cost can be improved to $(\log u, 1/\phi \log u)$, i.e. we do not require $\mathcal{V}$ to store the heavy hitter nodes. This is accomplished by having the prover, at each level of $\mathcal{T}$, "replay" the hash values of all nodes listed in the previous round. $\mathcal{V}$ can keep a simple fingerprint of the identities and hash values of all nodes listed in each round (computing their hash values internally), and compare this to a fingerprint of the hash values and identities listed by $\mathcal{P}$. If these fingerprints match for each level, $\mathcal{V}$ is assured that the correct information was presented. Note each node is repeated just once, so this only doubles the communication cost. This reduced cost protocol is used in Section 6.2.

**$k$-largest.** Given the same set up as the PREDECESSOR query, the $k$-th largest problem is to find the largest $p$ in the stream such that there are at least $k-1$ larger values $p'$ also present in the stream. This can be solved by the prover claiming that the $k$th largest item occurs at location $j$, and performing the range query protocol with the range $(j, u)$, allowing $\mathcal{V}$ to check that there are exactly $k$ distinct items present in the range. This has a cost of $(\log u, k + \log u)$. For large values of $k$, alternative approaches via range sum (assuming all keys are distinct) can reduce the cost to $(\log u, \log u)$.

### 6.2 Frequency-based Functions

Given the approach described in Section 3, it is natural to ask what other functions can be computed via sum-check protocols applied to carefully chosen polynomials. By extending the ideas from the protocol of Section 3, we get protocols for any statistic $F$ of the form $F(\mathbf{a}) = \sum_{i \in [u]} h(\mathbf{a}_i)$. Here, $h : \mathbb{N}_0 \to \mathbb{N}_0$ is a function over frequencies. Any statistic $F$ of this form is called a *frequency-based function*. Such functions occupy an important place in the streaming world. For example, setting $h(x) = x^2$ gives the self-join size. We will subsequently show that using functions of this form we can obtain non-trivial protocols for problems including:

- $F_0$, the number of distinct items in stream $A$.
- $F_{max}$, the frequency of the most-frequent item in $A$.
- Point queries on the inverse distribution of $A$. That is, for any $i$, we will obtain protocols for determining the number of tokens with frequency exactly $i$.

**The Protocol.** A natural first attempt to extend the protocols of Section 3 to this more general case is to have $\mathcal{V}$ compute $f_\mathbf{a}(\mathbf{r})$ as in Section 3, then have $\mathcal{P}$ send polynomials which are claimed to match sums over $h(f_\mathbf{a}(\mathbf{x}))$. In principal, this approach will work: for the $F_2$ protocol, this is essentially the outline with $h(x) = x^2$. However, recall that when this technique was generalized to $F_k$ for larger values of $k$, the cost increased with $k$. This is because the degree of the polynomial $h$ increased. In general, this approach yields a solution with cost $\deg(h) \log n$. This does not yet yield interesting results, since in general, the degree of $h$ can grow arbitrarily high, and the resulting protocol is worse than the trivial protocol which simply sends the entire vector **a** at a cost of $O(\min(n, u))$.

To overcome this obstacle, we modify this approach to use a polynomial function $\tilde{h}$ with bounded degree that is sublinear in $n$ and $u$. At a high level, we "remove" any very heavy elements from the stream $A$ before running the protocol of Section 3.1, with $f_\mathbf{a}^2$ replaced by $\tilde{h} \circ f_\mathbf{a}$ for a suitably chosen polynomial $\tilde{h}$. By removing all heavy elements from the stream, we keep the degree of $\tilde{h}$ (relatively) low, thereby controlling the communication cost. We now make this intuition precise.

Assume $n = O(u)$ and let $\phi = u^{-1/2}$. The first step is to identify the set $H$ of $\phi$-heavy hitters (i.e. the set of elements with frequency at least $u^{1/2}$) and their frequencies. We accomplish this via the $(\log u, 1/\phi \log u)$ protocol described in Section 6.1. $\mathcal{V}$ runs this protocol and, as the heavy hitters are reported, $\mathcal{V}$ incrementally computes $F' := \sum_{i \in H} h(\mathbf{a}_i)$, which can be understood as the contribution of the heaviest elements to $F$, the statistic of interest.

In parallel with the heavy hitters protocol, $\mathcal{V}$ also runs the first part of the protocol of Section 3.1 with $d = \log u$. That is, $\mathcal{V}$ chooses a random location $\mathbf{r} = (r_1, \ldots, r_d) \in [p]^d$ (where $p$ is a prime chosen larger than the maximum possible value of $F$), and while observing the stream $\mathcal{V}$ incrementally evaluates $f_\mathbf{a}(\mathbf{r})$. As in Sections 2 and 3.1, this requires only $O(d)$ additional words of memory.

As the heavy hitters are reported, $\mathcal{V}$ "removes" their contribution to $f_\mathbf{a}$ by subtracting $\mathbf{a_v}\chi_\mathbf{v}(\mathbf{r})$ from $f_\mathbf{a}(\mathbf{r})$ for each $\mathbf{v}\in H$. That is, let $\widetilde{f_\mathbf{a}}$ denote the polynomial implied by the derived stream obtained by removing all occurrences of all $\phi$-heavy hitters from $A$. Then $\mathcal{V}$ may compute $\widetilde{f_\mathbf{a}}(\mathbf{r})$ via the identity $\widetilde{f_\mathbf{a}}(\mathbf{r}) = f_\mathbf{a}(\mathbf{r}) - \sum_{\mathbf{v}\in H}\chi_\mathbf{v}(\mathbf{r})$. Crucially, $\mathcal{V}$ need not store the items in $H$ to compute this value; instead, $\mathcal{V}$ subtracts $\chi_\mathbf{v}(\mathbf{r})$ each time a heavy hitter $\mathbf{v}$ is reported, and then immediately "forgets" the identity of $\mathbf{v}$.

Now let $\tilde{h}$ be the unique polynomial of degree at most $u^{1/2}$ such that $\tilde{h}(i) = h(i)$ for $i = 0,\ldots,u^{1/2}$; $\mathcal{V}$ next computes $\tilde{h}(\widetilde{f_\mathbf{a}}(\mathbf{r}))$ in small space. Note that this computation can be performed without explicitly storing $\tilde{h}$, since we can compute

$$\tilde{h}(x) = \sum_{i=0,\ldots u^{1/2}} h(i)\chi_i(x)$$

(assuming $h()$ has a compact description as in the examples below).

The second part of the verification protocol can proceed in parallel with the first part. In the first round, the prover sends a polynomial $g_1(x_1)$ claimed to be

$$g_1(x_1) = \sum_{x_2,\ldots,x_d\in[\ell]^{d-1}} \tilde{h}\circ\widetilde{f_\mathbf{a}}(x_1,x_2,\ldots,x_d).$$

Observe that if $g_1$ is as claimed, then

$$F(\mathbf{a}) = \sum_{x_1\in[\ell]} g_1(x_1) + F' - |H|h(0).$$

Since the polynomial $g_1(x_1)$ has degree at most $u^{1/2}$, it can be described in $u^{1/2}$ words.

Then, as in Section 3.1, $\mathcal{V}$ sends $r_{j-1}$ to $\mathcal{P}$ in round $j > 1$. In return, the prover sends a polynomial $g_j(x_j)$, and claims

$$g_j(x_j) = \sum_{x_{j+1},\ldots,x_d\in[\ell]^{d-j}} \tilde{h}\circ\widetilde{f_\mathbf{a}}(r_1,\ldots,r_{j-1},x_j,x_{j+1},\ldots,x_d).$$

The verifier conducts tests for correctness that are completely analogous to those in Section 3.1, which completes the description of the protocol. The proof of completeness and soundness of this protocol is analogous to those in Section 3.1 as well.

**Analysis of space and communication.** $\mathcal{V}$ requires $\log u$ words to run the heavy hitters protocols, and $O(d) = O(\log u)$ space to store $r_1,\ldots,r_d$, $f_\mathbf{a}(\mathbf{r})$, $\widetilde{f_\mathbf{a}}(\mathbf{r})$, and to compute and store $\tilde{h}(\widetilde{f_\mathbf{a}}(\mathbf{r}))$. The communication cost of the heavy hitters protocol is $u^{1/2}\log u$, while the communication cost of the rest of the protocol is bounded by the $du^{1/2} = u^{1/2}\log u$ words used by $\mathcal{P}$ to send a polynomial of degree at most $u^{1/2}$ in each round. Thus, we have the following theorem:

THEOREM 6. *Assume $n = \Theta(u)$. There is a $\log u$ round, $(\log u, u^{1/2}\log u)$-protocol for any statistic $F$ of the form $F(\mathbf{a}) = \sum_{i\in[u]} h(\mathbf{a}_i)$, with probability of failure $O(\frac{\log u}{u})$. The verifier takes time $O(\log u)$ per update. The prover takes time $O(u^{3/2})$.*

Using this approach yields protocols for the following problems:

- $F_0$, the number of items with non-zero count. This follows by observing that $F_0$ is equivalent to computing $\sum_{i\in[u]} h(\mathbf{a}_i)$ for $h(0) = 0$ and $h(i) = 1$ for $i > 0$.

- More generally, we can compute functions on the inverse distribution, i.e. queries of the form "how many items occur exactly $k$ times in the stream" by setting, for any fixed $k$, $h(k) = 1$ and $h(i) = 0$ for $i \neq j$. One can build on this to compute, e.g. the number of items which occurred between $k$ and $k'$ times, the median of this distribution, etc.

- We obtain a protocol for $F_{max} = \max_i \mathbf{a}_i$, with a little more work. $\mathcal{P}$ first claims a lower bound $lb$ on $F_{max}$ by providing the index of an item with frequency $F_{max}$, which $\mathcal{V}$ verifies by running the INDEX protocol from Section 4. Then $\mathcal{V}$ runs the above protocol with $h(i) = 0$ for $i \leq lb$ and $h(i) = 1$ for $i > lb$; if $\sum_{i\in[u]} h(\mathbf{a}_i) = 0$, then $\mathcal{V}$ is convinced no item has frequency higher than $lb$, and concludes that $F_{max} = lb$.

COROLLARY 2. *There is a $(\log u, u^{1/2}\log u)$-protocol that requires just $\log u$ rounds of interaction for $F_0$, $F_{max}$, and queries on the inverse distribution.*

**Comparison.** Compared to the previous protocols, the methods above increase the amount of communication between the two parties by a $u^{\frac{1}{2}}$ factor. The number of rounds of interaction remains $\log u$, equivalent to $\mathcal{V}$'s space requirement. So arguably these bounds are still good from the verifier's perspective. In contrast, the construction of [14] requires $\Omega(\log^2 u)$ rounds of interaction and communication, which may be large enough to be offputting. To make this concrete, for a terabyte-size input, $\log u$ rounds is of the order of 40, while $\log^2 u$ is of the order of thousands. Meanwhile, the $u^{\frac{1}{2}}$ communication is of the order of a megabyte. So although the total communication cost is higher, one can easily imagine scenarios where the latency of network communications makes it more desirable to have fewer rounds with more communication in each.

## 7. CONCLUDING REMARKS

We have presented interactive proof protocols for various problems that are known to be hard in the streaming model. By delegating the hard computation task to a possibly dishonest prover, the verifier's space complexity is reduced to $O(\log u)$. We now outline directions for future study.

**Multiple Queries.** Many of the problems considered are parameterized by values that are only specified at query time. The results of these queries could cause the verifier to ask new queries with different parameters. However, re-running the protocols for a new query with the same choices of random numbers does not provide the same security guarantees. The guarantees rely on $\mathcal{P}$ not knowing these values; with this knowledge a dishonest prover could potentially find collisions under the polynomials, and fool the verifier.

Two simple solutions partially remedy this issue: firstly, it is safe to run multiple queries *in parallel* round-by-round using the same randomly chosen values, and obtain the same guarantees for each query. This can be thought of as a 'direct sum' result, and holds also for the Goldwasser *et al.* construction [14]. Secondly, $\mathcal{V}$ can just carry out multiple independent copies of the protocol. Since each copy requires only $O(\log u)$ space (more precisely $\log u + 1$ integers), the cost per query is low. Nevertheless, it remains of some practical interest to find protocols which can be used repeatedly to support an larger number of queries. Related work based on strong cryptographic assumptions has recently appeared [7, 12] but is currently impractical.

**Distributed Computation.** A motivation for studying this model arises from the case of cloud computation, which outsources computation to the more powerful "cloud". In practice, the cloud may in fact be a distributed cluster of machines, implementing a model such as Map-Reduce. We have so far assumed that the prover operates a traditional centralized computational entity. The next step is to study how to create proofs over large data in the distributed model. A first observation is that the proof protocols we give here naturally lend themselves to this setting: observe that the prover's message in each round can be written as the inner product of the input data with a function defined by the values of $r_j$ revealed so far. Thus, these protocols easily parallelize, and fit into Map-Reduce settings very naturally; it remains to demonstrate this empirically, and to establish similar results for other protocols.

**Other query types.** From a complexity perspective, the main open problem is to more precisely characterize the class of problems that are solvable in this streaming interactive proof model. We have shown how to modify the construction of [14] to obtain (poly log $u$, poly log $u$) streaming protocols for all of NC, and we showed that a wide class of reporting and aggregation queries possess (log $u$, log $u$) protocols. It is of interest to establish what other natural queries possess (log $u$, log $u$) protocols: $F_0$ and $F_{\max}$ are the prime candidates to resolve; other targets include other common queries, such as nearest neighbors. Determining whether problems outside NC possess interactive proofs (streaming or otherwise) with poly log $u$ communication and a verifier that runs in nearly linear time is a more challenging problem of considerable interest. This question asks, in essence, whether *parallelizable* computation is more easily verified than sequential computation.

## Acknowledgements

We thank Roy Luo for providing prototype protocol implementations. We also thank Michael Mitzenmacher, Salil Vadhan, Kai-Min Chung and Guy Rothblum for several helpful discussions.

## APPENDIX
## A. RESULTS DUE TO PRIOR WORK

**Streaming Universal Arguments.** A probabilistically checkable proof (PCP) is a proof in redundant form, such that the verifier need access only a few (randomly chosen) bits of the proof before deciding whether to accept or reject. A Universal Argument effectively simulates a PCP while ensuring $\mathcal{P}$ need not send the entire proof to $\mathcal{V}$. We first describe this simulation, before describing a particular PCP system that, when simulated by a Universal Argument, can be executed by a streaming verifier.

For a language $L$ on input $\mathbf{a}$, a Universal Argument consists of four messages: First, $\mathcal{V}$ sends $\mathcal{P}$ a collision-resistant hash function $h$. Next, an honest $\mathcal{P}$ constructs a PCP $\pi$ for $\mathbf{a}$, and then constructs a Merkle tree of $\pi$ using $h$ (the leaves of the tree are the bits of $\pi$) [20]. $\mathcal{P}$ then sends the value of the root of the tree to $\mathcal{V}$. This effectively "commits" $\mathcal{P}$ to the proof $\pi$; $\mathcal{P}$ cannot subsequently alter it without finding collisions for $h$. Third, $\mathcal{V}$ sends $\mathcal{P}$ a list of the locations of $\pi$ he needs to query. Finally, for each bit $b_i$ that is queried, $\mathcal{P}$ responds with the value of all nodes on $b_i$'s authentication path in the Merkle tree (note this path has only logarithmic length). $\mathcal{V}$ checks, for each bit $b_i$ that the authentication path is correct relative to the value of the root; if so, $\mathcal{V}$ is convinced $\mathcal{P}$ returned the correct value for $b_i$ as long as $\mathcal{P}$ cannot find a collision for $h$. The theorem follows by combining this construction with the fact that there exist PCP systems in which $\mathcal{V}$ only needs access to $\mathbf{a}$ in order to evaluate $O(1)$ locations in the LDE $f_{\mathbf{a}}$. We now justify this last claim by describing such a PCP system.

In [5], Ben-Sasson et al. describe for any language in NP a PCP system in which $\mathcal{V}$ is not given explicit access to the input; instead, $\mathcal{V}$ has oracle access to an encoding of the input $\mathbf{a}$ under an *arbitrary* error-correcting code (to simplify a little). In their PCP system, $\mathcal{V}$ runs in polylogarithmic time and queries only $O(1)$ bits of the encoded input, and $O(1)$ bits of the proof $\pi$. Moreover, these bits are determined non-adaptively (specifically, they do not depend on $\mathbf{a}$). We show this implies a PCP system that satisfies the claim for any $L \in$ NP. Indeed, let $LDE(\mathbf{a})$ denote the truth-table of $f_{\mathbf{a}}$; i.e. $LDE(\mathbf{a})$ is a list of elements in the field $\mathbb{Z}_p$, one for each $\mathbf{r} \in \mathbb{Z}_p^d$. There are (two-stage) *concatenated codes* whose first stage applies the $LDE$ operation to the input $\mathbf{a}$ (and whose second stage applies a code to turn the field elements in $LDE(\mathbf{a})$ into bits) that suffice as encodings of $\mathbf{a}$ [2]. Therefore, a streaming verifier with explicit access to the input $\mathbf{a}$ may simulate the verifier $\mathcal{V}$ in the PCP system of Ben-Sasson et al: each time $\mathcal{V}$ queries a bit $b_i$ of the encoded input, there is a location $\mathbf{r}$ such that $b_i$ can be extracted from $f_{\mathbf{a}}(\mathbf{r})$.

A Universal Argument based on the PCP of the previous paragraph has two additional properties worth mentioning. First, since $\mathcal{V}$ need only query $O(1)$ bits of $f_{\mathbf{a}}$ and otherwise runs in poly log time, we obtain a streaming verifier that runs in *near-linear* time. Second, since $\mathcal{V}$ need only query $O(1)$ bits of the proof, and the authentication path of each bit in the Merkle tree is of length $O(\log u)$, the communication cost of the Universal Argument is $O(\log u)$ words. Putting all these pieces together yields Theorem 2.

**Streaming "Interactive Proofs for Muggles."**[2] In [14], $\mathcal{V}$ and $\mathcal{P}$ first agree on a circuit $C$ of fan-in 2 that computes the function of interest; $C$ is assumed to be in layered form. $\mathcal{P}$ begins by claiming a value for the output gate of the circuit. The protocol then proceeds iteratively from the output layer of $C$ to the input layer, with one iteration for each layer. Let $\mathbf{v}^{(i)}$ be the vector of values that the gates in $i$-th layer of $C$ take on input $x$, with layer 1 corresponding to the output layer, and let $f_{\mathbf{v}^{(i)}}$ be the LDE of $\mathbf{v}^{(i)}$.

At a high level, in iteration 1, $\mathcal{V}$ reduces verifying the claimed value of the output gate to verifying $f_{\mathbf{v}^{(2)}}(\mathbf{r})$ for a random location $\mathbf{r}$. Likewise, in iteration $i$, $\mathcal{V}$ reduces verifying $f_{\mathbf{v}^{(i)}}$ to verifying $f_{\mathbf{v}^{(i+1)}}(\mathbf{r}')$ for a random $\mathbf{r}'$. Critically, the verifier's final test requires only $f_{\mathbf{v}^{(d)}}(\mathbf{r}) = f_{\mathbf{a}}(\mathbf{r})$, the low-degree extension of the input at the random location $\mathbf{r}$, which can be chosen at random independent of the data or the circuit, and hence computed by a streaming verifier. Note that each iteration takes logarithmically many rounds, with a constant number of words of communication in each round. Therefore the protocol requires $O(d \log u)$ communication in total. In particular, all problems that can be solved in log-space by non-streaming algorithms (i.e. algorithms that can make multiple passes over the input) possess polynomial size circuits of depth $\log^2 u$, and hence there are $(\log^3 u, \log^3 u)$ protocols for these problems.

## B. DETAILED PROOFS
### B.1 Analysis of SELF-JOIN SIZE

**Proof of correctness.** We now argue in detail that the verifier is unlikely to be fooled by a dishonest prover.

LEMMA 1. *If the prover follows the above protocol then the verifier will accept with certainty. However, if the prover sends any polynomial which does not meet the required property, then the verifier will accept with probability at most $2dl/p$, where this probability is over the random coin tosses of $\mathcal{V}$.*

PROOF. The first part is immediate from the following discussion: if each $g_j$ is as claimed, then the verifier can easily ensure that each $g_j$ is consistent with $g_{j-1}$.

For the second part, the proof proceeds from the $d$th round back to the first round. In the final round, the prover has sent $g_d$, of degree $2\ell - 2$, and the verifier checks that it agrees with a precomputed value at $x_d = r_d$. This is an instance of the Schwartz-Zippel polynomial equality testing procedure [24]. If $g_d$ is indeed as claimed, then the test will always be passed, no matter what the value of $r_d$. But if $g_d$ does not satisfy the equality, then $\Pr[g_d(r_d) = f^2(\mathbf{r})] \leq \frac{2\ell-2}{p}$. Therefore, if $p$ was chosen so that $p \gg \ell$, then the verifier is unlikely to be fooled.

---

[2]This result was observed by Guy Rothblum; here, we present the details of the construction for completeness.

The argument proceeds inductively. Suppose that the verifier is convinced (with some small probability of error) that $g_{j+1}(x_{j+1})$ is indeed as claimed, and wants to be sure that $g_j(x_j)$ is also as claimed. The prover has claimed that

$$g_j(x_j) = \sum_{x_{j+1},\ldots,x_d \in [\ell]^{d-j}} f_{\mathbf{a}}^2(r_1,\ldots,r_{j-1},x_j,x_{j+1},\ldots,x_d).$$

We again verify this by a Schwartz-Zippel polynomial test: we evaluate $g_j(x_j)$ at a randomly chosen point $r_j$, and ensure that the result is correct. Observe that

$$g_j(r_j) = \sum_{x_{j+1},\ldots,x_d \in [\ell]^{d-j}} f_{\mathbf{a}}^2(r_1,\ldots,r_j,x_{j+1},\ldots x_d)$$
$$= \sum_{x_{j+1} \in [\ell]} \sum_{x_{j+2},\ldots,x_d \in [\ell]^{d-j-1}} f_{\mathbf{a}}^2(r_1,\ldots,r_j,x_{j+1},x_{j+2},\ldots,x_d)$$
$$= \sum_{x_{j+1} \in [\ell]} g_{j+1}(x_{j+1}).$$

Therefore, if the verifier $\mathcal{V}$ believes that $g_{j+1}$ is as claimed, then (provided the test passes) $\mathcal{V}$ has enough confidence to believe that $g_j$ is also as claimed. More formally,

$$\Pr\left[g_j \neq \sum_{x_{j+1},\ldots,x_d \in [\ell]^{d-j}} f_{\mathbf{a}}^2(r_1,\ldots r_{j-1},x_j,\ldots,x_d) \,\middle|\, g_{j+1} \equiv \sum_{x_{j+2},\ldots,x_d \in [\ell]^{d-j+1}} f_{\mathbf{a}}^2(r_1,\ldots r_j,x_{j+1},\ldots,x_d)\right] < \frac{2\ell}{p}.$$

In the final step, the verifier is satisfied that $g_1$ is consistent with $g_2$, and so $g_1$ is as claimed. The probability that $g_1$ is not as claimed can be bounded as the probability that the verifier was fooled in any intervening step. This is at most $2d\ell/p$, by a union bound.

Intuitively, the key reason for the prover's inability to fool the verifier is that the prover must *commit* to a particular $g_j$ before $r_j$ is revealed to him. So while the prover could then choose a $g_{j+1}$ which causes the test on that pair to pass, $g_{j+1}$ is also "dishonest". But ultimately, the prover must provide $g_d$, which $\mathcal{V}$ can check based on information that is known to $\mathcal{V}$ alone. The prover is very unlikely to have included a dishonest $g_j$ along the way and passed all the subsequent tests to generate a $g_d$ which is consistent with the final test using $r_d$ (which remains unknown to $\mathcal{P}$). □

**Analysis of prover's costs.** Besides the verifier's space and communication, this protocol is also quite efficient in terms of the other costs. Let us fix $\ell = 2$. As the stream is being processed the verifier has to update the LDE $f_{\mathbf{a}}(\mathbf{r})$. The updates are very simple, since $\chi_0(x) = 1 - x$ and $\chi_1(x) = x$, so

$$\chi_{\mathbf{v}}(\mathbf{r}) = \prod_{j=1}^{d}((1-v_j)(1-r_j) + v_j r_j).$$

Thus processing each update in the stream $O(d) = O(\log u)$ time.

The prover has to retain the input vector $\mathbf{a}$, which can be done efficiently in space $O(\min(u,n))$. In the verification process it is clear that the verifier spends $O(1)$ time per round evaluating a degree-2 polynomial, so the total time is $O(\log u)$. On the prover side, it might appear costly to compute each $g_j(x_j)$ naively following the definition. But observe that $g_j(x_j)$ is a polynomial of degree 2, so it is sufficient to evaluate $g_j(x_j)$ at three locations, say at $x_j = 0, 1, 2$, to determine $g_j(x_j)$. For a location $x_j = c$, we rewrite

$$g_j(c) = \sum_{x_{j+1},\ldots,x_d \in [\ell]^{d-j}} f_{\mathbf{a}}^2(r_1,\ldots,r_{j-1},c,x_{j+1},\ldots x_d)$$

$$= \sum_{x_{j+1},\ldots,x_d \in [\ell]^{d-j}} \left(\sum_{\mathbf{v} \in [\ell]^d} a_{\mathbf{v}} \chi_{\mathbf{v}}(r_1,\ldots,r_{j-1},c,x_{j+1},\ldots x_d)\right)^2$$
$$= \sum_{x_{j+1},\ldots,x_d \in [\ell]^{d-j}} \sum_{\mathbf{v_1},\mathbf{v_2} \in [\ell]^d} a_{\mathbf{v_1}} a_{\mathbf{v_2}} \chi_{\mathbf{v_1}}(r_1,\ldots,r_{j-1},c,x_{j+1},\ldots x_d)$$
$$\cdot \chi_{\mathbf{v_2}}(r_1,\ldots,r_{j-1},c,x_{j+1},\ldots x_d)$$
$$= \sum_{\mathbf{v_1},\mathbf{v_2} \in [\ell]^d} \left(a_{\mathbf{v_1}} a_{\mathbf{v_2}} \prod_{k=1}^{j-1} \chi_{v_{1,k}}(r_k) \cdot \chi_{v_{1,j}}(c) \cdot \prod_{k=1}^{j-1} \chi_{v_{2,k}}(r_k) \cdot \chi_{v_{2,j}}(c)\right.$$
$$\left.\cdot \sum_{x_{j+1}\ldots x_d \in [\ell]^{d-j}} \left(\prod_{k=j+1}^{d} \chi_{v_{1,k}}(x_k)\chi_{v_{2,k}}(x_k)\right)\right).$$

Note that $\chi_{v_k}(x_k) = 1$ iff $x_k = v_k$ and 0 for any other value in $[\ell]$, for any pair of $\mathbf{v_1}, \mathbf{v_2}$, we have

$$\sum_{x_{j+1},\ldots,x_d \in [\ell]^{d-j}} \left(\prod_{k=j+1}^{d} \chi_{v_{1,k}}(x_k)\chi_{v_{2,k}}(x_k)\right) = 1$$

if and only if $\forall j+1 \leq k \leq d : v_{1,k} = v_{2,k}$, and 0 otherwise. Thus,

$$g_j(c) = \sum_{\mathbf{v_1},\mathbf{v_2} \in [\ell]^d, \forall j+1 \leq k \leq d: v_{1,k}=v_{2,k}} \left(a_{\mathbf{v_1}} a_{\mathbf{v_2}} \prod_{k=1}^{j-1} \chi_{v_{1,k}}(r_k)\right.$$
$$\left.\cdot \chi_{v_{1,j}}(c) \prod_{k=1}^{j-1} \chi_{v_{2,k}}(r_k)\chi_{v_{2,j}}(c)\right)$$
$$= \sum_{v_{j+1},\ldots,v_d \in [\ell]^{d-j}} \left(\sum_{v_1,\ldots,v_j \in [\ell]^j} \left(a_{\mathbf{v}}\chi_{v_j}(c)\prod_{k=1}^{j-1} \chi_{v_k}(r_k)\right)\right)^2.$$

$\mathcal{P}$ maintains $a_{\mathbf{v}} \prod_{k=1}^{j-1} \chi_{v_k}(r_k)$ for each nonzero $a_{\mathbf{v}}$, updating with the new $r_k$ in each round as it is revealed in constant time. Thus the total time spent by the prover for the verification process can be bounded via $O(n \log u)$, where $n$ is the number of nonzero $a_{\mathbf{v}}$'s.

We make one further simplification. At the heart of the computation is a summation over $[\ell]^j$ for each $v_{j+1},\ldots,v_d \in [\ell]^{d-j}$. As we set $\ell = 2$,

$$\sum_{v_1,\ldots,v_j \in [\ell]^j} \left(a_{\mathbf{v}}\chi_{v_j}(c)\prod_{k=1}^{j-1} \chi_{v_k}(r_k)\right)$$
$$= \sum_{v_j=0}^{1} \left(\chi_{v_j}(c) \cdot \sum_{v_1,\ldots,v_{j-1} \in [\ell]^{j-1}} \left(a_{\mathbf{v}} \prod_{k=1}^{j-1} \chi_{v_k}(r_k)\right)\right)$$

And for each $v_j,\ldots,v_d \in [\ell]^{d-j+1}$, we can decompose

$$\sum_{v_1,\ldots,v_{j-1} \in [\ell]^{j-1}} \left(a_{\mathbf{v}} \prod_{k=1}^{j-1} \chi_{v_k}(r_k)\right)$$
$$= \sum_{v_{j-1}=0}^{1} \left(\chi_{v_{j-1}}(r_{j-1}) \sum_{v_1,\ldots,v_{j-2} \in [\ell]^{j-2}} \left(a_{\mathbf{v}} \prod_{k=1}^{j-2} \chi_{v_k}(r_k)\right)\right).$$

By storing $A_j[v_j \ldots v_d] = \sum_{v_1 \ldots v_{j-1} \in [\ell]^{j-1}} (\mathbf{a_v} \prod_{k=1}^{j-1} \chi_{v_k}(r_k))$, $\mathcal{P}$ computes $A_{j+1}[v_{j+1} \ldots v_d] = \chi_0(r_j) A_j[0, v_{j+1} \ldots v_d] + \chi_1(r_j) A_j[1, v_{j+1} \ldots v_d]$ in time $O(u/2^j)$. The total time is $O(\min(n \log(u/n), u))$, at most linear in $u$. Note that computing the $F_2$ alone takes $\Theta(\min(n,u))$ time, so there is at most a logarithmic factor more work than simply providing the answer.

## B.2 Analysis of SUB-VECTOR Protocol

PROOF OF THEOREM 5. **Correctness.** It is clear that with an honest $\mathcal{P}$, $\mathcal{V}$ always accepts. Next, we argue that if $\mathcal{P}$ returns a wrong value in any round, then $t' \neq t$ with high probability. $\mathcal{P}$ first sends back $a'_i$ for all $q_L \leq i \leq q_R$ and their siblings (if they are outside of the range). Consider any pair of siblings, say $a'_i$ and $a'_{i+1}$. Consider the functions $f(x) = a_i + a_{i+1}x$ and $f'(x) = a'_i + a'_{i+1}x$ in the field $\mathbb{Z}_p$. If $a_i \neq a'_i$ or $a_{i+1} \neq a'_{i+1}$, the two linear functions will not be identical, and they will intersect at no more than one point in $[p]$. Since we choose $r_1$ uniformly randomly from $[p]$, the probability that $f(r_1) = f'(r_1)$ is at most $1/p$. Thus, if $\mathcal{P}$'s first message is not correct, with probability at least $1 - 1/p$, there will be at least one error in the computed $\gamma^{(1)}(i)$, $q_L \leq i \leq q_R$. The same argument applies to each of the following $(\log u - 1)$ rounds: if either of the siblings of $\gamma^{(j)}(q_L)$ and $\gamma^{(j)}(q_R)$ returned by $\mathcal{V}$ is wrong or some $\gamma^{(j)}(i), q_L \leq i \leq q_R$ is already wrong previously, then with probability at most $1/p$, the reconstructed $\gamma^{(j)}(i)$ will be all correct. By the union bound, the probability that an incorrect response from $\mathcal{V}$ will lead to a correct $t'$ is at most $\frac{\log u}{p}$.

**Analysis of costs.** We first argue that $\mathcal{V}$ can compute $t$ in small space. Expanding $t$, we have

$$t = \sum_i \Big( a_i \prod_{j=1}^{\log u} r_j^{(i-1)_j} \Big), \qquad (8)$$

where $(i-1)_j$ denotes the $j$-th least significant bit of the binary representation of $i-1$. Initially when $\mathbf{a} = \mathbf{0}$, we have $t = 0$; when we have $a_i \leftarrow a_i + \delta$, $t$ is incremented (modulo $p$) by $\Delta t = \delta \cdot \prod_{j=1}^{\log u} r_j^{(i-1)_j}$, which is easily computed in $O(\log u)$ time. Thus $\mathcal{V}$ can maintain $t$ by just keeping $t, r_1, \ldots, r_{\log u}$.

The verifier's space requirement for the protocol is also bounded by $O(\log u)$ words. Given the query range, as the sub-vector result arrives at $\mathcal{V}$, the verifier can keep track of only $O(\log u)$ hash values of internal nodes, corresponding to at most one child of $\gamma^j(q_L)$ and $\gamma^j(q_R)$ for each $j$. Combining these with the hash values provided by $\mathcal{P}$ will be sufficient to run the checking protocol. Each of these can be maintained in small space in the same manner as the root $t$ via (8) above. Thus the space to carry out the protocol is $O(\log u)$.

The communication cost consists of the initial query result of size $k$ sent by the prover, plus $O(1)$ nodes per level of the binary tree $\mathcal{T}$. So the total communication cost is $O(\log u + k)$.

Now we analyze the prover's cost. As the stream is received the prover clearly needs linear space and $O(1)$ time per element to construct the vector $\mathbf{a}$. At verification time the prover essentially reconstructs the binary tree $\mathcal{T}$. Note that $\mathcal{T}$ has at most $n$ nonzero leaves, so it has size $O(\min(u, n \log(u/n)))$. Computing this tree in a bottom-up fashion costs $O(1)$ time per node, hence $O(\min(u, n \log(u/n)))$ time in total. □

*Remarks.* As in Section 3 the failure probability can be driven down to $O(\frac{\log u}{u^c})$ for any constant $c$ by picking $p$ greater than $u^c$, without changing the asymptotic bounds. From the description above a dishonest prover may cause excessive communication by sending more than $k$ nonzero entries in the initial answer. To guarantee the $O(\log u + k)$ bound with any $\mathcal{P}$, we could first verify the value of $k$, i.e., a RANGE-COUNT query, with $O(\log u)$ communication using the protocol in Section 3. Then if $\mathcal{P}$ sends more than $k$ nonzero entries $\mathcal{V}$ will reject immediately.

We note that by modifying the hash function to $(1 - r_j)v_L + r_j v_R$, it is possible to show that $t$ is equivalent to the LDE $f(\mathbf{r})$, while the same analysis holds. This provides a connection between the two approaches, although the proofs are quite different in nature.

## C. REFERENCES

[1] N. Alon, Y. Matias, and M. Szegedy. The space complexity of approximating the frequency moments. *J. Comp Sys Sci*, 58:137–147, 1999.

[2] S. Arora and B. Barak. *Computational Complexity: A Modern Approach*. Cambridge University Press, 2009.

[3] S. Arora and S. Safra. Probabilistic checking of proofs: A new characterization of NP. *J. ACM*, 45(1):70–122, 1998.

[4] L. Babai and S. Moran. Arthur-merlin games: a randomized proof system, and a hierarchy of complexity class. *J. Comput. Syst. Sci.*, 36(2):254–276, 1988.

[5] E. Ben-Sasson, O. Goldreich, P. Harsha, M. Sudan, and S. Vadhan. Short PCPs verifiable in polylogarithmic time. In *CCC*, pages 120–134, 2005.

[6] A. Chakrabarti, G. Cormode, and A. McGregor. Annotations in data streams. In *ICALP*, pages 222–234, 2009.

[7] K.-M. Chung, Y. Kalai, and S. Vadhan. Improved delegation of computation using fully homomorphic encryption. In *CRYPTO*, pages 483–501, 2010.

[8] G. Cormode and S. Muthukrishnan. An improved data stream summary: The count-min sketch and its applications. *J. Algorithms*, 55(1):58–75, 2005.

[9] G. DeCandia *et al.* Dynamo: Amazon's Highly Available Key-value Store. In *SOSP*, pages 205–220, 2007.

[10] P. Flajolet and G. N. Martin. Probabilistic counting algorithms for data base applications. *J. of Comp Sys Sci*, 31(2):182–209, 1985.

[11] L. Fortnow and C. Lund. Interactive proof systems and alternating time-space complexity. *Theoretical Computer Science*, 113(1):55–73, 1993.

[12] R. Gennaro, C. Gentry, and B. Parno. Non-interactive verifiable computing: Outsourcing computation to untrusted workers. In *CRYPTO*, pages 465–482, 2010.

[13] S. Garfinkel. Sub-Linear Drive Analysis. http://simson.net/page/Sub-Linear_Drive_Analysis

[14] S. Goldwasser, Y. T. Kalai, and G. N. Rothblum. Delegating computation: Interactive proofs for Muggles. In *STOC*, pages 113-122, 2008.

[15] S. Goldwasser and M. Sipser. Private coins versus public coins in interactive proof systems. In *STOC*, pages 59–68, 1986.

[16] A. Juels and B. Kaliski. PORs: Proofs of retrievability for large files. In *Computer and Communications Security*, pages 584–597, 2007.

[17] J. Kilian. A note on efficient zero-knowledge proofs and arguments (extended abstract). In *STOC*, pages 723–732, 1992.

[18] E. Kushilevitz and N. Nisan. *Communication Complexity*. Cambridge University Press, 1997.

[19] F. Li, K. Yi, M. Hadjieleftheriou, and G. Kollios. Proof-infused streams: Enabling authentication of sliding window queries on streams. In *VLDB*, pages 147–158, 2007.

[20] R. Merkle. *Secrecy, authentication, and public key systems*. PhD thesis, Electrical Engineering, Stanford, 1979.

[21] S. Muthukrishnan. *Data streams: algorithms and applications*. *Found. Trends Theor. Comput. Sci.*, 2005.

[22] S. Papadopoulos, Y. Yang, and D. Papadias. Continuous authentication on relational streams. *VLDB J.*, 19(2):161–180, 2010

[23] A. D. Sarma, R. J. Lipton, and D. Nanongkai. Best-order streaming model. In *TAMC*, pages 178-191, 2009.

[24] J. Schwartz. Fast probabilistic algorithms for verification of polynomial identities. *J. ACM*, 27(4):701–717, 1980.

[25] A. Shamir. IP = PSPACE. *J. ACM*, 39(4):869–877, 1992.

[26] S. Setty, A. J. Blumberg, and M. Walfish. Toward practical and unconditional verification of remote computations. In *Proc. of HotOS Workshop*, pages 1–5, 2011.

[27] Y. Yang, S. Papadopoulos, D. Papadias, and G. Kollios. Authenticated indexing for outsourced spatial databases. *VLDB J.*, 18(3):631–648, 2009.

[28] K. Yi, F. Li, G. Cormode, M. Hadjieleftheriou, G. Kollios, and D. Srivastava. Small synopses for group-by query verification on outsourced data streams. *ACM Transactions on Database Systems*, 34(3), article 15, 2009.